\documentclass[twocolumn,aps,prb,superscriptaddress,amssymb,showpacs,relax]{revtex4-1}
\usepackage[dvips,final]{graphicx}
\begin{document}

\title{Comment on ``Dirac cones reshaped by interaction effects in suspended graphene"
by Elias et al., Nature Physics 7, 701 (2011)}

\author{N. Garc\'ia}\email{nicolas.garcia@fsp.csic.es}
\affiliation{Laboratorio de F\'isica de Sistemas Peque\~nos y
Nanotecnolog\'ia,
 Consejo Superior de Investigaciones Cient\'ificas, E-28006 Madrid, Spain}
 \affiliation{Laborat\'orio de Filmes Finos e Superficies, UFSC,
 88.040-900 Florianopolis, Brazil}
\author{P. Esquinazi}\email{esquin@physik.uni-leipzig.de}
\affiliation{Division of Superconductivity and Magnetism, Institut
f\"ur Experimentelle Physik II, Universit\"{a}t Leipzig,
Linn\'{e}stra{\ss}e 5, D-04103 Leipzig, Germany}
\author{A. A.. Pasa}
\affiliation{Laborat\'orio de Filmes Finos e Superficies, UFSC,
 88.040-900 Florianopolis, Brazil}
 
\maketitle

In a recent paper Elias et al.\cite{eli11} studied the reshaping
of the Dirac cone in graphene due to electron-electron interaction
using renormalization group theory and compared it to measurements
of the velocity $v(n)$ when the carrier concentration $n
\rightarrow 0$. The small parameter of expansion was the inverse
of the fermion species in graphene $1/N_F$, with $N_F= 4$. The
result they found is that the velocity diverges logarithmically as
the energy or the density of carriers approaches zero. The
velocity becomes then:
\begin{equation}
    v(n) = v_F(n_0) [1 + b \ln(n_0/n)]\,,
    \end{equation}
where $n_0$  is the concentration that corresponds to the
ultraviolet cut-off energy and $v_F(n_0)$ is the Fermi velocity
near the cut-off. This result is not new and was obtained more
than ten years ago \cite{gon94,gon99} and it is not for $N_F = 4$
but for $N_F \rightarrow \infty$. However, it is expected that for
$N_F = 4$ it is a good approximation, although when $N_F$
decreases the result is poorer, see Refs.~\onlinecite{fos08,son07}
for this discussion. Son \cite{son07} proposed that $N_F = 2$, one
for the spin. Let us now consider that Eq.~(1) is a good
approximation. Then the dispersion relation is:
\begin{equation}
E(k) = k(1 - a \ln(k))\,,
    \end{equation}
where $a$ is a constant and $k$ the wavevector measured from the
$H$ point of the Brillouin zone. The question that arises now is
where are the Dirac carriers? The dispersion relation is not
anymore linear. Are there Dirac fermions or not?

Figure~1 shows the dispersion relations for electrons with
positive, zero and negative masses. It can be noticed that
electrons with a dispersion relation given by Eq.~(2) have
actually negative mass. In this case they are actually holes for
energy larger than zero and the holes for energy smaller than zero
should be electrons. Indeed, not a really clear, simple
description for the band carriers.
\begin{figure}[]
\begin{center}
\includegraphics[width=1\columnwidth]{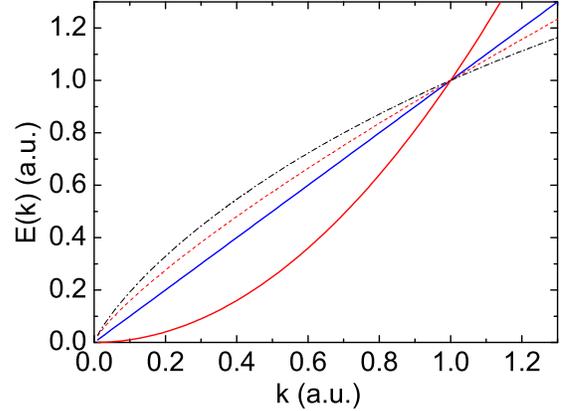}
\caption[]{Dispersion relation $E(k)$ vs. wavevector $k$ for: a
quadratic dispersion relation for positive electron mass (read
continous curve) $E(k) = k^2$, a linear dispersion relation $E(k)
= k$  for zero electronic mass (straight blue curve), a power law
relation $E(k) = k^{0.8}$ (dotted line) as proposed in
Ref.~\onlinecite{son07} and $E(k) = k(1 - 0.4 \ln(k))$ (black
dashed-dotted line). Note that the last two dispersion relations
imply negative masses for the carriers.} \label{nT-MG}
\end{center}
\end{figure}

In addition, the bands are difficult to discriminate between
Eq.~(2) or a law like $k^{1 - \alpha}$ that may correspond to
small $N_F$. In Fig.~1  we have plotted this dispersion relation
because this is precisely what appears for $N_F$ finite, where
$\alpha \simeq 0.15 \ldots 0.2$, see Ref.~\onlinecite{son07}. By
choosing different parameters it is difficult to conclude, which
of the laws, the logarithmic or the power law, is the one that
would fit the data. We note that the authors in
Ref.~\onlinecite{eli11} insinuate that their data show an
excellent agreement with the used model for $N_F$ large. But
actually the scattering of the data is rather large and the
agreement is  poor.

The approximated theory for small $N_F$ predicts that for $N_F <
2.53$ a band gap appears in the excitation spectrum, something
that would be difficult to get from the approximations used by the
authors in Ref.~\onlinecite{eli11}. However, this has been solved
later by Drut and L{\"a}hde \cite{druprl09} using Monte Carlo
simulations and proved that a band gap appears for the case of
ideal graphene and small $N_F$. The existence of a narrow gap has
been recently shown by transport measurements in mesoscopic, thin
graphite samples \cite{gar11}. The existence of an energy gap will
solve automatically the apparent difficulties between holes and
electrons discussed before and the dispersion at low energies will
be rather parabolic or other complicated behavior but not linear.

In conclusion, two different points of view are available to
understand the behavior of graphene at low energies. One is
considering a large $N_F$,  that makes graphene a semimetal, and
another for small $N_F < 2.5$  that would make graphene a narrow
gap semiconductor\cite{son07}, a prediction supported
independently by Monte Carlo simulations \cite{druprl09}. Taking
into account recently obtained experimental evidence for weakly
coupled graphene layers, i.e. graphite \cite{gar11}, we tend to
support the last one.



%

\end{document}